\documentclass[12pt,preprint]{iopart}
\usepackage{graphicx,enumerate}
\usepackage{amssymb}
\usepackage{epsfig}

\begin{document}

\title{Complex Network Analysis of Brazilian Power Grid}

\author{G. C. Martins$^{1}$, L. S. Oliveira$^{1}$, F. L. Ribeiro$^{2}$ and F. L. Forgerini$^{1}$}

\address{$^{1}$ Campus Paulo Freire, Universidade Federal do Sul da Bahia \\
45988-058 \hspace{5mm} Teixeira de Freitas - BA \hspace{5mm} Brazil\\

$^{2}$ Departamento de F\'isica, Universidade Federal de Lavras \\
37200-000, Caixa Postal 3037 \hspace{5mm} Lavras - MG \hspace{5mm} Brazil}

\ead{fabricio.forgerini@ufsb.edu.br}

\begin{abstract}
\noindent
Power Grids and other delivery networks has been attracted some attention by the network literature last decades. Despite the Power Grids dynamics has been controlled by computer systems and human operators, the static features of this type of network can be studied and analyzed. The topology of the Brazilian Power Grid (BPG) was studied in this work. We obtained the spatial structure of the BPG from the ONS (electric system’s national operator), consisting of high-voltage transmission lines, generating stations and substations. The local low-voltage substations and local power delivery as well the dynamic features of the network were neglected. We analyze the complex network of the BPG and identify the main topological information, such as the mean degree, the degree distribution, the network size and the clustering coefficient to caracterize the complex network. We also detected the critical locations on the network and, therefore, the more susceptible points to lead to a cascading failure and even to a blackouts. Surprisely, due to the characteristic of the topology and physical structure of the network, we show that the BPG is resilient against random failures, since the random removal of links does not affect significantly the size of the largest cluster. We observe that when a fraction ρ of the links are randomly removed, the network may disintegrates into smaller and disconnected parts, however, the power grid largest component remains connected. We believe that the even a static study of the network topology can help to identify the critical situations and also prevent failures and possible blackouts on the network.
\end{abstract}

\maketitle

\section{Introduction}
\label{intro}

Power Grids are part of what have been called technological networks, the physical insfrastructure network for the transportation of energy and information~\cite{newman_book}. The Internet and telephone networks are some examples of these type of networks. This field has received some attention from the researches and the correspondent literature last decades mainly due to the problems caused by the possibility of large scale failures leading to major blackouts~\cite{sergey_book, forgerini_book, motter, dobson}. Despite the fact that the Power Grids dynamics has been controlled by computer systems and human operators~\cite{Watts}, the static features of this type of networks can be studied and provide valuable information for private or public investments and prevention of failures~\cite{arianos}.

The topology of the Brazilian Power Grid (BPG) was studied in this work. We obtained the spatial structure of the BPG from the Brazilian electric system's operator, ONS, consisting of the high voltage transmission lines, generating stations and substations. The local low-voltage substations were neglected in this work. We also neglected the dynamic features of the power grid, such as the electromagnetic processes and human interference by controlling the energy flux on the network.

We analyze the BPG as a complex network in which all the stations and substations were represented by nodes and the transmission lines were represented by links among the nodes. By identifying the main topological information and use these features to find out the network's weakness, one can give indications for improvements of the BPG in order to prevent failures and blackouts.

This paper is organized as follows. In Sec.~2 we present our Methodology and the details for data analysis. In Sec.~3 we describe our main results and discussions. Finally, in Sec.~4, we summarize the results and present our main conclusions.


\section{Methodology}
\label{Methodology}

The BPG data was obtained from the ONS database, regarding to the 2015 update. The network analysis was performed by using the methodos of networks theory, focusing on some topological properties, namely the mean degree and the degree distribution, betweenness centrality, transitivity and the connectivity of the network.

The degree $k_i$, is the number of links attached to a node $i$, in a total of $N$ nodes, given by the number of transmission lines connected to the station or substation. On the other hand, the mean degree of the network is a global measure, given by

\begin{equation}
 \langle k_i \rangle = \frac{1}{N} \sum_{i=1}^N k_i.
\end{equation}

The degree distribution $P(k)$ is the probability that a node chosen uniformly at random within the network has a degree $k$:

\begin{equation}
\label{equation_dd}
 P(k) = \frac{\langle N(k) \rangle}{N},
\end{equation}
where $N$ is the total number of nodes in the network and $\langle N(k) \rangle$ is the average number (or fraction) of nodes with degree $k$ on the network~\cite{forgerini_book}.

The distribution is given by the equation~\ref{equation_dd} and, once it is known, much information can be obtained by the calculation of moments of this distribution. The $n$-th moment of the distribution is

\begin{equation}
 \langle k^n \rangle = \sum_{k=0}^\infty k^n P(k).
\end{equation}
The first moment $\langle k \rangle$ is the mean degree, while the second moment is a mesure of the degree fluctuations of the distribution. If $\langle k^2 \rangle$ diverges, structure and function dramatically changes in the network, in contrast to those for finite $\langle k \rangle$~\cite{boccaletti}.

An exponential degree distribution has the form
\begin{equation}
P(k) = Ce^{-\alpha k},
\end{equation}
where $C$ is an arbitrary constant. Networks with exponential degree distributions were reported in some real-world situations such as the Power Grids~\cite{albert}.

\begin{figure}[!htb]
    \vskip 0.2cm
    \begin{minipage}[t]{0.48 \linewidth}
	\includegraphics[width=\linewidth]{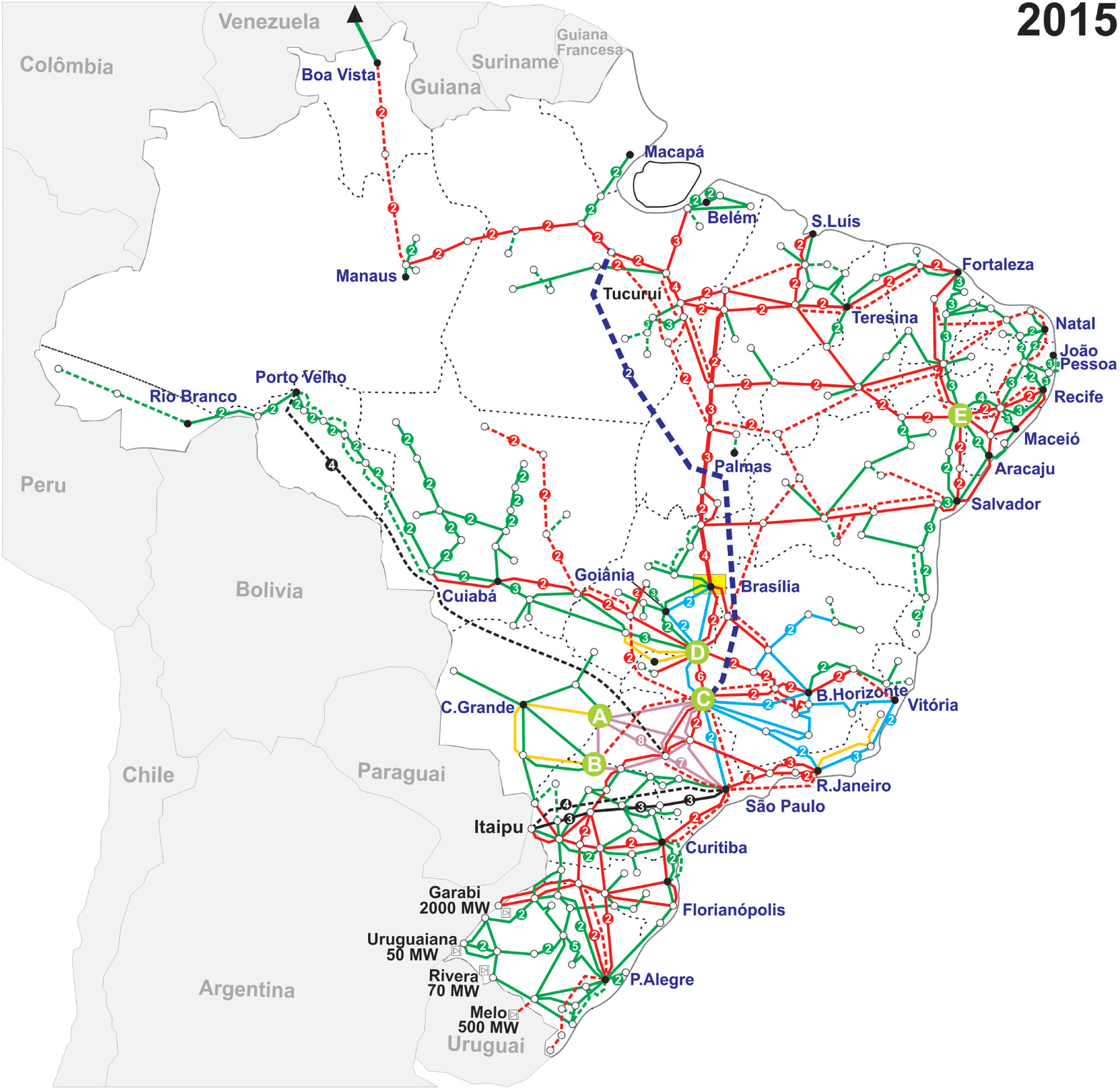}\\
    \end{minipage}\hfill
    \begin{minipage}[t]{0.48 \linewidth}
	\includegraphics[width=\linewidth]{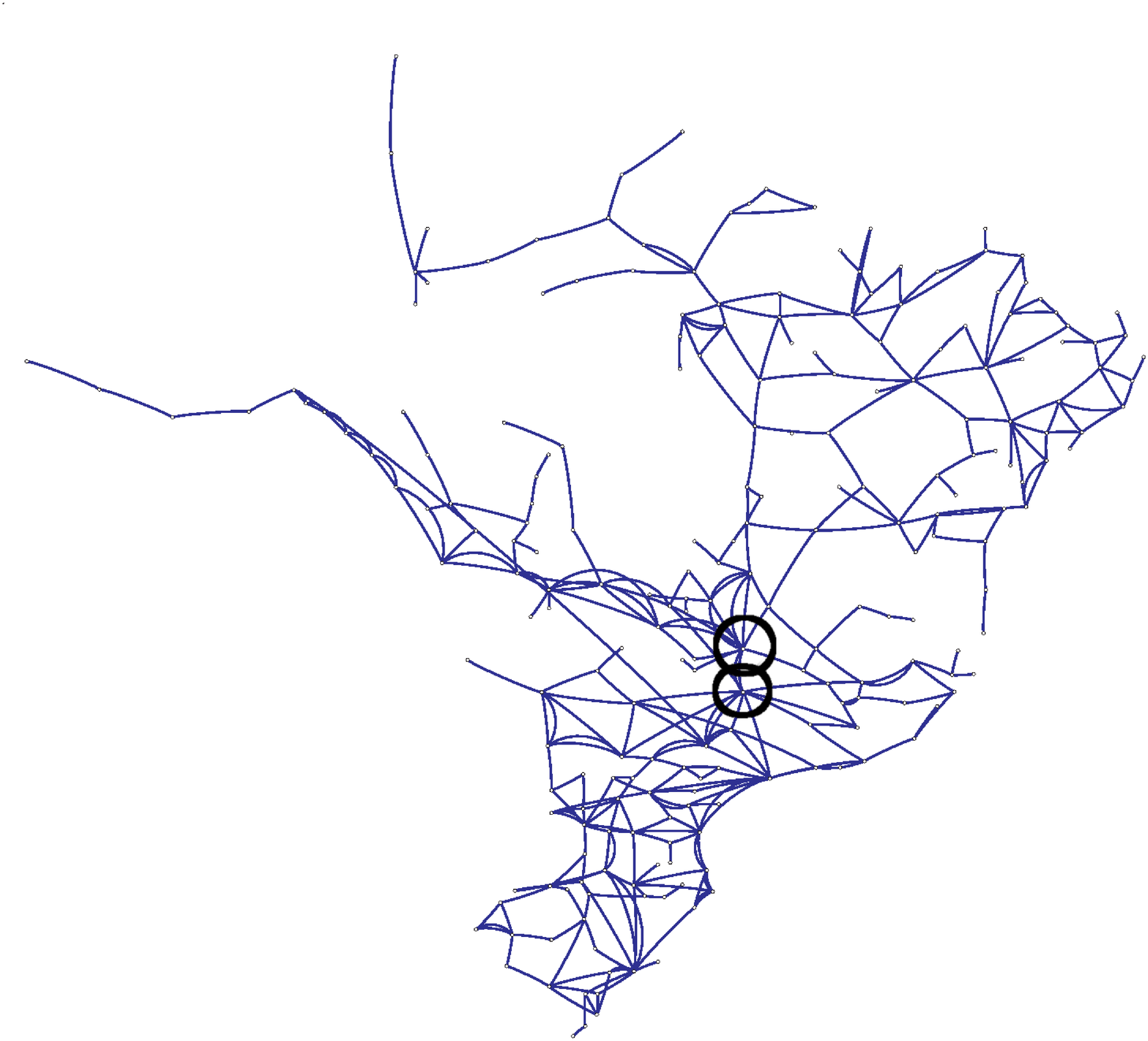}\\
    \end{minipage}\hfill
 \caption{The Brazilian Power Grid (left panel) and this network representation (right panel). The traced lines at the left panel represent the transmission lines under construction and were not take into account in this work as one can see at the right panel. We also show, inside the black circles, very important nodes playing the role of keep the network connected, as will be discussed in section~\ref{Results}.}
 \label{fig1}
\end{figure}

We also can visualize the network structure with the corresponding positions to the generating stations and substations and the transmission lines connecting them. As one can see in the Fig.~\ref{fig1}, we show the Brazilian Power Grid and it's corresponding complex network. This picture was created by using the large network analysis software Pajek$\texttrademark$\cite{pajek}.

As mentioned before, in this work we do not take into account the electromagnetic processes, computer controlling and human interference in the energy flux of the network. Despite the fact that we neglect the dynamic features of the network, our analysis perhaps it is not so naive after all. As we will show in next section, the BPG is strongly dependent of the topological structure. In case of some large structural failures, the network can be lead to cascading failures that may not be corrected by human interference or computer systems. We believe that the this approach can provide powerful information about the BPG's vulnerability.


\section{Results and Discussion}
\label{Results}

The BPG has evolved greatly in the last decades but so far does not show itself as a complex as other power grids worldwide, in comparison to countries with approximately same size or population than Brazil. Only for sake of comparison, the USA's power grid is represented by a network of 14099 nodes and 19657 links among them~\cite{albert}.

The main results of the characterization of the BPG are shown in the table 1. One can see that these measures represent the most basic topological information of the network.

\begin{table}[!htb]
\label{table}
\centering
\caption{Main results of the topological analysis of the Brazilian Power Grid and, for sake of comparison, the values of the US Power Grid~\cite{Watts, albert}.}
\footnotesize
\begin{tabular}{@{}lcc}
\br
Characteristic & Numerical Value (Brazil) & Numerical Value (US) \\
\mr
Mean Degree & 6.487 & 1.394\\
Number of nodes & 230 & 14099\\
Number of links & 1492 & 19657\\
Clustering coefficient & 0.174 & 0.080\\
Betweenness Centrality & 0.438 & -\\
\br
\end{tabular}\\

\end{table}

The degree distribution is a very important network measure. Much information about a network is related to degree distribution. Networks with heavy tail scale free degree distribution have been investigated over a decade.  A network with a power-law degree distribution, $P(k) \sim k^{-\gamma}$ with $2 \leq \gamma \le 3$, for instance, is expected to be resilient to a random removal of links, \textit{i.e.} resilient against random failures of the grid~\cite{cohen_havlin_2}. Numerous real world networks, including the Internet, have a power-law degree distribuition with $\gamma \approx 3$ and in all of them were observed the resilience phenomena~\cite{barabasi}. Networks may have the degree distribution with exponent even greater the 3. For the case where $\gamma > 4$, this type of networks are kwnown as ultra-resilient against random failures. 

\begin{figure}[!htb]
\centering
  \includegraphics[width=0.6\linewidth]{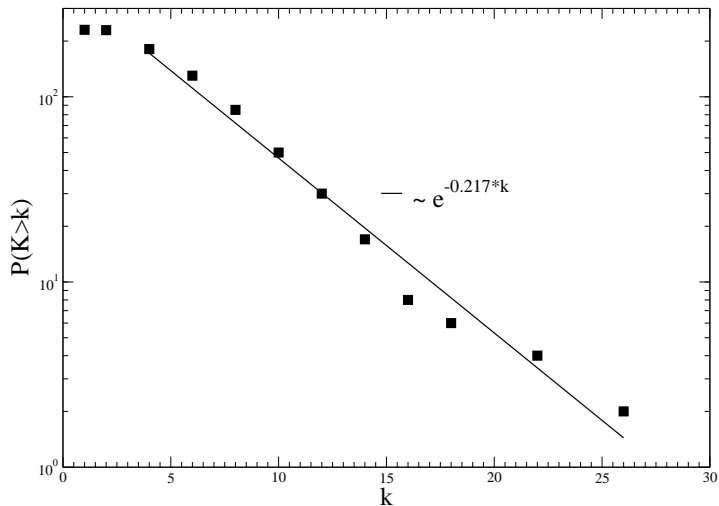}
\caption{Cumulative degree distribution, $P(K>k)$, of the Brazilian Power Grid. As shown in the picture of semilogarithmic scale, the degree distribution has an exponential form.}
 \label{fig2}
\end{figure}

In this work we use the cumulative degree distribution,
\begin{equation}
 P(K>k) = \sum_{k=K}^N P(k),
\end{equation}
which is the probability that the degree is greater or equal to k. By using the cumulative degree distribution to represent the data, we reduce the noise in the tail of the distribution. As one can see in Fig.~\ref{fig2}, the cumulative degree distribution is shown and it has the exponential form. 

Another important measure is the mean clustering coefficient of the network, $\overline{c}$. The concept of clustering reflects how the first neighbors of a node are connected to each other. This feature is a statistic of the number of multiple connections among the nodes in the network. In other words, clustering coefficient refers to the statistics of the number of triangles (loops of length 3) in the network. The clustering coefficient of the entire network, or the \textit{mean clustering coefficient}, is the average of the local clustering coefficient over all nodes:
\begin{equation}
 \overline{c} = \sum_k P(k)\overline{c}(k),
\end{equation}
and may vary between 0 and 1. In the BPG we found $\overline{c} = 0.174$, a very small value for the clustering coefficient. This indicate that the BPG is, in average, poorly connected and highly susceptible to the formation of isolated clusters in the network.

In order to investigate the connectivity of the network and how it should be affected by random removal of links and nodes we studied the number of connected components on the network, $C$ and also the size of the largest cluster on the network, $S$. As one can see by our simulations, both measurements of connectivity presents different and complementary results.

\begin{figure}[!htb]
\vspace{0.55cm}
\centering
  \includegraphics[width=0.6\linewidth]{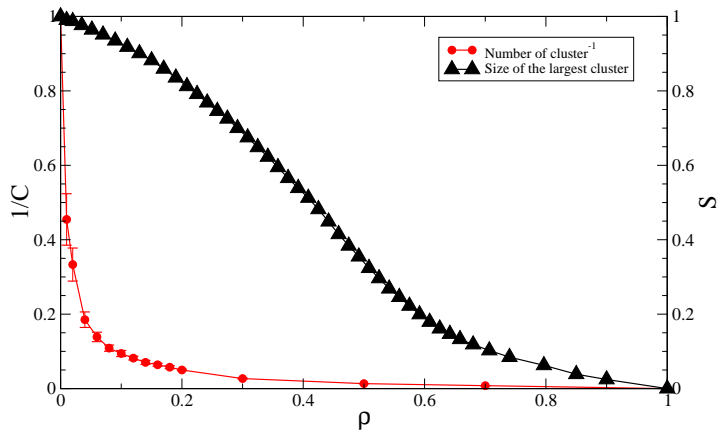}
\caption{Mesurements of the network connectivity as function of the fraction of nodes removed at random from the network. We measure the number of connected components on the network, $C$ and the size of the largest cluster on the network, $S$. In the plot we show on the left axis the 1/$C$ versus $\rho$ and, on the rigth axis, S versus $\rho$. The results were obtained averaging 100 different realizations. This result suggest that even the network turn into a set of disconnected parts, for small $\rho$, the Brazilian Power Grid is resilient against random failures.}
 \label{fig3}
\end{figure}

For the first measurement, when all the links are present, the fraction of removed links, $\rho$, is equal zero and the network is a single connected component. For the case, $1/C = 1$. As the number of removed links increases, the network is divided into disconnected parts and $1/C \rightarrow 0$. When the fraction of removed links, $\rho = 1$, $1/C = 0$. We performed simulations to investigate the network connectivity. By randomly removal of links from the network, we calculate the number of connected components on the newtork, for different realizations. Our results are shown in the figure~\ref{fig3}, where one can see on the left axis the plot 1/$C$ versus $\rho$. We show that for a small fraction of links removed, $\rho \approx 0.1$, the Brazilian Powed Grid turn into a set of several disconnected parts (small value of $1/C$). The power grid rupture is significantly important for the Brazilian case, since the number of generating stations is relatively small.

The second connectivity measurement that were studied is $S$. As one can see on the right axis of the figure~\ref{fig3}, the decaing of $S$ is not as fast as $1/C$. For small values of $\rho$, $S$ is almost linear, following by an exponential regime. As shown, even for small values of $\rho$, the network turn into a set of several disconnected parts. However, most of these disconnected parts are isolated nodes, not affecting substantially the connected component of the network. Only for larger values of $\rho$ the size of largest cluster $S$ shows a rapidly decreasing. This result show that the topological structure of the Brazilian Power Grid prove itself, in some sense, resilient against random failues. 

Analyzing the BPG one can see that there are few nodes playing an important role by keeping the network connected. The most important (and susceptible) nodes are shown in figure~\ref{fig1} as two black circles. These points consists in large substation and hidrelectric power plants, called \textit{Furnas Centrais El\'etricas}. These critical locations on the network make the connection between the south-southeast and the center of the country and then to the north and northeast regions. The importance of the nodes are due to the failures in these major electric complex will lead the BPG system to a cascading failure and even to a major blackouts.


\section{Conclusions}
\label{Conclusions}

In this work we studied and analyzed the topology of the Brazilian Power Grid (BPG). Despite the fact that the BPG is comparatively much smaller than power grids from other countries, by analyzing its spatial structure, we show that the BPG is poorly connected and highly dependent on a few nodes to maintenance the network's connected component. We calculated and summarize most basic topological information of the BPG.

The degree distribution of the network has an exponential form and, according to the newtork literature, are not resilient against random failures. We also simulate random failures on the network by removing links at random. Our results show that with a small fraction of missing connections ($\rho \approx 0.1$), the network turns into a set of several disconnected parts, leading parts of the country isolated. However, these isolated parts of the network does not represent the larger part of the power grid. We show that the random removing of links does not affect significantly, for small values of $\rho$, the size of the largest cluster on the network, $S$, keeping the most of the country connected by power grid.

In conclusion, even with a static analysis of the network, one can see the main topological information and it's weakness. To improve the network and become more resilient against random failues, BPG need some improvements regarding to the number of connections to prevent failures and future blackouts.


\section*{Acknowledgments}

The authors would like to thank the Federal University of Southern Bahia, UFSB, for the financial support by project 005/2015-PROSIS.

\end{document}